\documentclass[twocolumn,showpacs,preprintnumbers,amsmath,amssymb,prl,superscriptaddress]{revtex4}

\usepackage{graphicx,here}
\usepackage{dcolumn}
\usepackage{bm}
\usepackage{enumerate}

\begin{document}

\preprint{APS/123-QED}

\title{Piezo-Magnetoelectric Effect of Spin Origin in Dysprosium Orthoferrite}

\author{Taro Nakajima}
\email{taro.nakajima@riken.jp}
\affiliation{RIKEN Center for Emergent Matter Science (CEMS), Saitama 351-0198, Japan.}
\author{Yusuke Tokunaga}
\affiliation{RIKEN Center for Emergent Matter Science (CEMS), Saitama 351-0198, Japan.}
\affiliation{Department of Advanced Materials Science, University of Tokyo, Kashiwa 277-8561, Japan}
\author{Yasujiro Taguchi}
\affiliation{RIKEN Center for Emergent Matter Science (CEMS), Saitama 351-0198, Japan.}
\author{Yoshinori Tokura}
\affiliation{RIKEN Center for Emergent Matter Science (CEMS), Saitama 351-0198, Japan.}
\affiliation{Department of Applied Physics and Quantum-Phase Electronics Center (QPEC), University of Tokyo, Tokyo 113-8656, Japan}
\author{Taka-hisa Arima}
\affiliation{RIKEN Center for Emergent Matter Science (CEMS), Saitama 351-0198, Japan.}
\affiliation{Department of Advanced Materials Science, University of Tokyo, Kashiwa 277-8561, Japan}

\begin{abstract}
Piezo-magnetoelectric effect, namely simultaneous induction of both the ferromagnetic moment and electric polarization by an application of uniaxial stress, was achieved in the non-ferroelectric and antiferromagnetic ground state of DyFeO$_3$.
The induced electric polarization and ferromagnetic moment are coupled with each other, and monotonically increase with increasing uniaxial stress. %
The present work provides a new way to design spin-driven multiferroic states, that is, magnetic symmetry breaking forced by external uniaxial stress.
\end{abstract}

\maketitle

The discovery of the spin-driven ferroelectricity in TbMnO$_3$\cite{Kimura_nature} opened the door to new frontiers in multiferroics, in which different kinds of ferroic orders coexist \cite{Fiebig,Arima_SpinDriven_Review,Tokura_MF_Review}. %
The extensive studies on recently-explored multiferroics, such as orthorhombic $R$MnO$_3$ ($R$ is a rare-earth element) \cite{Kimura_nature,Kimura_RMnO3,Kenzelmann_PRL_Spiral}, Ni$_3$V$_2$O$_8$ \cite{PRL_Ni3V2O8}, MnWO$_4$ \cite{PRL_MnWO4} etc., demonstrated that ferroelectricity can arise from the inversion symmetry breaking due to magnetic orders. %
As for the microscopic mechanisms to explain the spin-driven ferroelectricity, the most robust scheme is the inverse Dzyaloshinskii-Moriya (IDM) mechanism \cite{inverse_DM}, in which the magnetically induced electric polarization ($P$) is described as $P\propto e_{i,j}\times(S_i\times S_j)$ \cite{Katsura_PRL_2005,Mostovoy_PRL_2006}, where $e_{i,j}$ is a unit vector connecting between two spins, $S_i$ and $S_j$. %
This formula predicts ferroelectricity in a `cycloidal' spin order, in which the spontaneous electric polarization appears along the twofold axis perpendicular to both the propagation wave vector and $(S_i\times S_j)$, regardless of the symmetry of the underlying chemical lattice. %

On the other hand, there are two other mechanisms for the spin-driven ferroelectricity \cite{Jia2,Arima_dp,Magnetostriction1,Magnetostriction2}, namely, the spin-dependent $p$-$d$ hybridization model \cite{Arima_dp} and the magnetostriction model \cite{Magnetostriction1,Magnetostriction2}. %
In these models, the emergence of the ferroelectricity cannot be explained by local spin arrangement alone, but is explained by taking into account the symmetry of the crystal structure. %
This suggests that in the $non$-IDM type spin-driven multiferroics, the ferroelectricity can be controlled via changes in crystal structural symmetry. %

One of the most primitive ways to control the symmetry of the crystal structure is an application of anisotropic stress. %
It has recently been demonstrated that an application of compressive uniaxial stress can induce spin-driven electric polarization in Ba$_2$CoGe$_2$O$_7$ \cite{Nakajima_BCGO_PRL}, in which the origin of the ferroelectricity is explained by the $p$-$d$ hybridization model \cite{PRL_Murakawa_BCGO}. %
However, this system originally has a non-centrosymmetric (but non-polar) crystal structure belonging to the space group of $P\bar{4}2_1m$, in which the application of uniaxial stress naturally leads to piezoelectric polarization regardless of the spin degree of freedom. %
Therefore, the uniaxial-stress-induced ferroelectricity in the antiferromagnetic phase might be trivial. %

In the present study, we have investigated uniaxial-stress effects on magnetic and dielectric properties of a dysprosium orthoferrite, DyFeO$_3$, which has a centrosymmetric crystal structure. %
This system is known to display multiferroic nature in a magnetic-field-induced phase above $\sim 3$ T \cite{Tokunaga_DFO_PRL}. %
The multiferroicity can be described in terms of the magnetostriction between the magnetic moments on Fe and Dy atoms \cite{Tokunaga_DFO_PRL,NPhys_Tokunaga_RFO}. %
On the other hand, the ground state is an antiferromagnetic and non-ferroelectric state. %
In this Letter, we report that an application of uniaxial stress induces ferroelectricity in the ground state. 
Moreover, the uniaxial-stress-induced ferroelectricity is accompanied by (weak) ferromagnetism. %
To the best of our knowledge, this is the first experimental observation of `piezo-magnetoelectric effect', namely simultaneous induction of both the electric polarization and ferromagnetic moment, in a single-phase compound. %
The present study also reveals that the induced ferromagnetic moment and electric polarization are tightly coupled with each other, indicating that the piezo-magnetoelectric responses are of spin origin. %

\begin{figure}[t]
\begin{center}
	\includegraphics[clip,keepaspectratio,width=8cm]{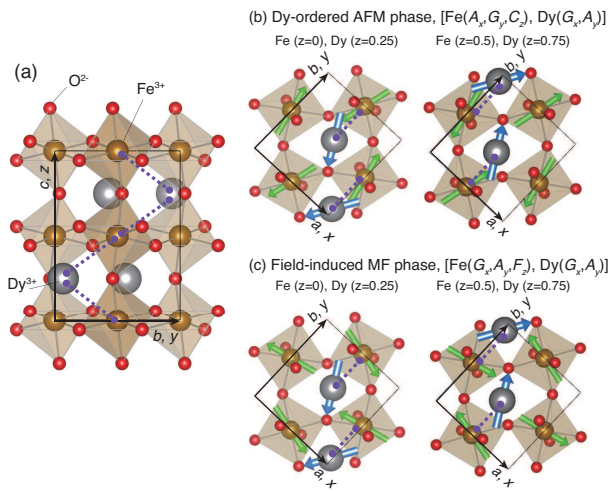}
	\caption{(Color online) (a) Crystal structure of DyFeO$_3$. Purple dotted lines show the paths of the nearest neighbor exchange coupling between the Fe spins and Dy moments. %
	Magnetic structures of (b) the Dy-ordered AFM phase and (c) the field-induced multiferroic phase. }
	\label{crystal}
\end{center}
\end{figure}

DyFeO$_3$ has an orthorhombically-distorted perovskite structure, as shown in Fig. \ref{crystal}(b). %
In this Letter, we employed a conventional $Pbnm$ (orthorhombic) setting. %
The magnetic property of this system has been investigated since 1960s \cite{DFO_Gorodetsky_JAP,RFO_Gorodetsky_PR}. %
At room temperature, the system exhibits the most conventional-type ($G$-type) antiferromagnetic order, in which the staggered Fe$^{3+}$ spins direct nearly along the $a$ axis, while the magnetic moments of Dy$^{3+}$ ions are disordered. %
Because of the DM interaction between the Fe spins, the spins are slightly canted from the $a$ axis, yielding a parasitic layered ($A$-type) antiferromagnetic and weak ferromagnetic components along the $b$ and $c$ axes, respectively \cite{RFO_Gorodetsky_PR}. %
According to the Bertaut's notation \cite{Bertaut}, this magnetic structure is described as $G_xA_yF_z$. %
We refer to this phase as the weak-ferromagnetic (WFM) phase. %

With decreasing temperature in zero magnetic field, the system undergoes a magnetic phase transition from the WFM phase to an antiferromagnetic (AFM) phase around $T_{\rm SR}=40$ K. %
This transition was identified to be a reorientation of the majority $G$-component from along the $a$ axis to the $b$ axis \cite{DFO_Gorodetsky_JAP}. %
This spin-reorientation leads to disappearance of the WFM component, and instead, the canted spin components, namely the $a$ and $c$ components exhibit the $A$-type and $C$-type orders, respectively. %
As a result, the magnetic order is described as $A_xG_yC_z$. %
This indicates that the direction of the $G$-component is relevant to the weak ferromagnetism in this system. %

With further decreasing temperature, Dy moments also exhibit an antiferromagnetic order below $T_{\rm N}^{\rm Dy}=4$ K. %
We refer to this phase as the Dy-ordered AFM phase. %
According to a previous magnetization study on isostructural DyAlO$_3$ \cite{PRB_DyAlO3}, the configuration of the Dy moments was reported to be $G_xA_y$. %
In Fig. \ref{crystal}(c), we show the spin arrangements in the Dy-ordered AFM phase. %
When a magnetic field is applied along the $c$ axis in this phase, the system exhibit a magnetic-field-induced phase transition, at which the configuration of the Fe spins turns back to $G_xA_yF_z$, around $H_{\rm SR}\sim3$ T. %
Tokunaga \textit{et al.} have reported that the electric polarization appears along the $c$ axis in the field-induced phase \cite{Tokunaga_DFO_PRL}. %
They explained the origin of the multiferroic nature by adapting the magnetostriction model to the nearest neighbor exchange coupling between the Fe spins and Dy moments, as shown in Fig. \ref{crystal}(c) \cite{Tokunaga_DFO_PRL,NPhys_Tokunaga_RFO}. %

Piezoelectric and piezomagnetic responses for these magnetic phases can be predicted in terms of the magnetic point group theory \cite{DFO_Gorodetsky_JAP}. %
For example, the magnetic point group for the AFM phase and the Dy-ordered AFM phase are `$mmm$' and `$222$', respectively. %
By applying uniaxial stress along the (110) direction in the AFM phase, we can break two mirrors out of the three, and therefore the magnetic point group is reduced to `$2/m$', which arrows ferromagnetic moment to appear along the $c$ axis. %
Similarly, in the Dy-ordered AFM phase, an application of uniaxial stress along the (110) direction reduces the magnetic point group symmetry down to `$2$', leading to ferroelectric polarization as well as ferromagnetic moment parallel to the $c$ axis. %
To verify these symmetry arguments, we performed magnetization and pyroelectric measurements under applied uniaxial stress. %

A single crystal of DyFeO$_3$ was grown by the floating-zone method and cut into a rectangular shape with dimensions of $2.1\times 2.3 \times 0.9$ mm$^3$. %
The widest surfaces were normal to the $c$ axis, and two of the other surfaces were selected to be the (110) plane. %
The magnetization measurements were performed using Magnetic Property Measurement System (Quantum Design Inc.) with the uniaxial-stress insert used in the previous studies \cite{Nakajima_BCGO_PRL,CFO_UniaxialPressPT}. %
For this experiment, we also developed a horizontal-uniaxial-stress cell, which is essentially the same as that used in the previous pyroelectric measurements on Ba$_2$CoGe$_2$O$_7$ \cite{Nakajima_BCGO_PRL}. %
The uniaxial compressive stress $\sigma$ was applied on the $(110)$ surfaces. %
Magnetic fields $H$ were applied along the $c$ axis. %
For the pyroelectric measurements, silver-paste electrodes were formed on the surfaces normal to the $c$ axis in order to observe the electric polarization along the $c$ axis ($P_c$). %
We also employed the uniaxial-stress insert for Physical Property Measurement System (Quantum Design Inc.), which was used in the previous studies \cite{Nakajima_BCGO_PRL,CFGO_2ax,CFO_UniaxialPressPT}. %
We measured displacement electric current with varying temperature $T$ or $\sigma$ using an electrometer (Keithley 6517B). %
By integrating the current with respect to time, we deduced $P_c$. %
Note that the magnitudes of $\sigma$ were calculated from the load applied from the top of the insert and the pressurized area of the sample. %

\begin{figure}[t]
\begin{center}
	\includegraphics[clip,keepaspectratio,width=7.5cm]{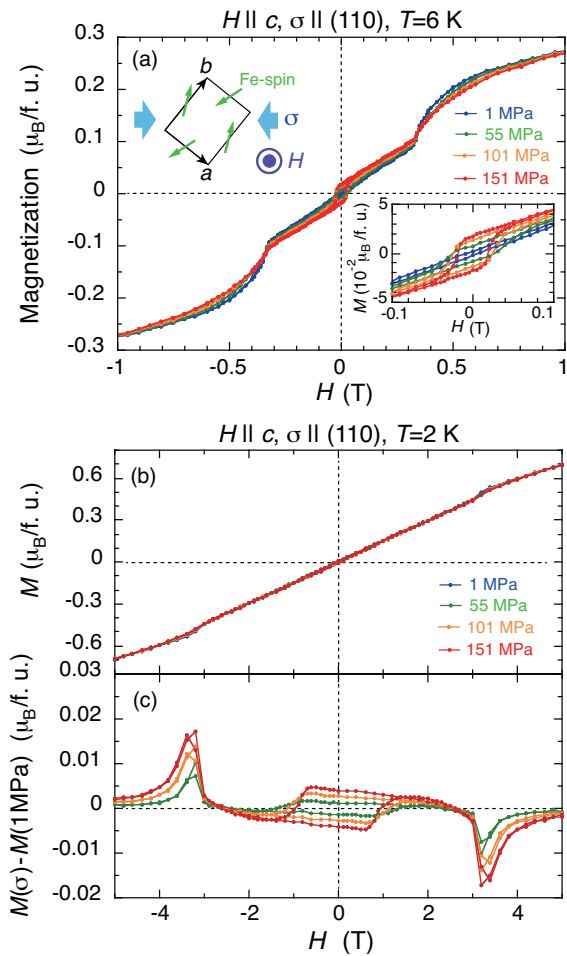}
	\caption{(Color online) (a) $M$-$H$ curves observed at 6 K under applied $\sigma$ of up to 151 MPa. %
	The insets show (left-top) a schematic showing the directions of the Fe spins in the AFM phase, $\sigma$ and $H$, and (right-bottom) a magnification of the hysteresis loops near $H=0$. %
	(b) $M$-$H$ curves and (c) their differences from the data at $\sigma=1$ MPa measured at 2 K under applied $\sigma$ of up to 151 MPa.  }
	\label{mag}
\end{center}
\end{figure}

Figure \ref{mag}(a) shows the results of the magnetization measurements at 6 K in the AFM phase. %
We observed that the $M$-$H$ curve shows a distinct hysteresis around $H=0$ under $\sigma$, revealing that the application of $\sigma$ induces weak ferromagnetism in the AFM phase. %
On the other hand, the critical field between the AFM phase to the field-induced WFM phase is hardly affected by the application of $\sigma$. %

We also measured magnetization curves at 2 K under $\sigma$, as shown in Fig. \ref{mag}(b). %
Although the changes in the $M$-$H$ curve were rather small, we succeeded in observing the ferromagnetic hysteresis loops under $\sigma$ by subtracting the data measured at $\sigma=1$ MPa from the other data, as shown in Fig. \ref{mag}(c). %
The remanent magnetization and coercive field increases and decreases with increasing $\sigma$, respectively. %

\begin{figure}[t]
\begin{center}
	\includegraphics[clip,keepaspectratio,width=8cm]{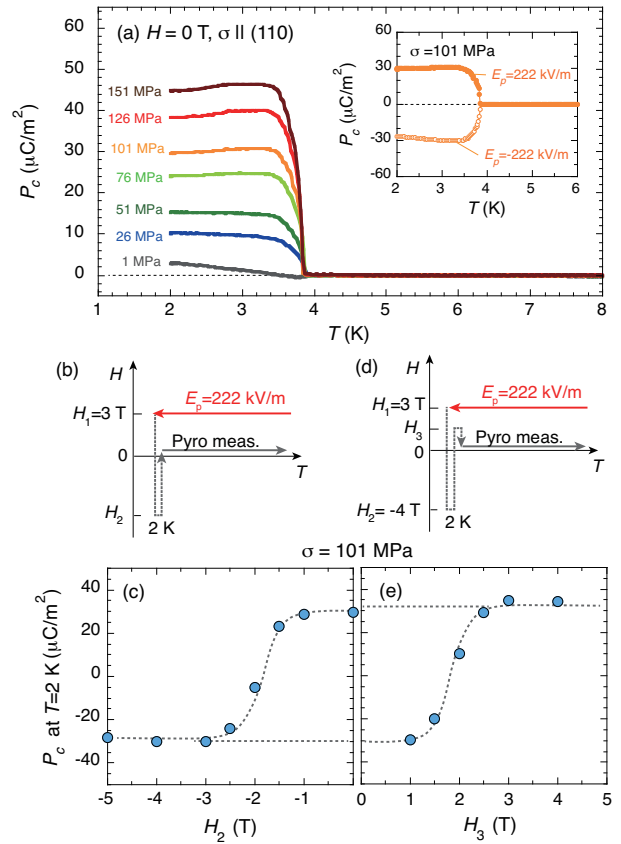}
	\caption{(Color online) (a) Temperature dependence of $P_c$ measured on heating in zero magnetic field after cooling under applied $\sigma$, $H=3$ T and $E_p=222$ kV/m. %
	The inset shows the data measured at $\sigma=101$ MPa after cooling with positive and negative poling electric fields. %
	(c) and (e) show the values of $P_c$ at 2 K measured by pyroelectric measurements in zero magnetic field after temperature and magnetic field sequences shown in (b) and (d), respectively. %
	}
	\label{pyro}
\end{center}
\end{figure}

Figure \ref{pyro}(a) shows the results of pyroelectric measurements. %
Before these measurements, the sample was cooled under applied $\sigma$, $H=3$ T and a poling electric field ($E_p$) of 222 kV/m, and then $H$ and $E_p$ were removed at 2 K. %
Subsequently, we measured pyroelectric current on heating under $\sigma$. %
We found that the application of $\sigma$ induces spontaneous electric polarization along the $c$ axis below $T_{\rm N}^{\rm Dy}$, %
and that the magnitude of $P_c$ increases with increasing $\sigma$. %
We also observed that the sign of $P_c$ was reversed when the sample was cooled in a negative $E_p$, as shown in the inset of Fig. \ref{pyro}(a). %
It is worth mentioning here that in the previous study on Ba$_2$CoGe$_2$O$_7$ \cite{Nakajima_BCGO_PRL}, polarity of the piezoelectric polarization was determined only by direction of the uniaxial stress. %
By contrast, in DyFeO$_3$, the application of $\sigma$ induces a genuine `ferroelectric' state, in which the spontaneous electric polarization can be reversed by an application of electric field.  %

To confirm the that $\sigma$-induced ferroelectricity is of spin origin, we investigated $H$ dependence of $P_c$ as follows. %
We cooled the sample under applied $\sigma=101$ MPa, $H=3$ T and $E_p=222$ kV/m. %
After removing $E_p$ at 2 K, we changed $H$ from 3 T to a negative magnetic field of $H_2$, and then removed it, as shown in Fig. \ref{pyro}(b). %
Figure \ref{pyro}(c) shows $H_2$ dependence of the values of $P_c$ at 2 K measured by the pyroelectric measurements after the above mentioned sequences. %
We found that the $\sigma$-induced $P_c$ was flipped by the application of $H_2$ below $\sim -2$ T. %
We also performed similar magnetic field sweeping sequences shown in Fig. \ref{pyro}(d), revealing that the flipped electric polarization was reversed again by applying a positive magnetic field of $H_3$ above $\sim 2$ T, as shown in Fig. \ref{pyro}(e). %
These results show that $P_c$ shows hysteresis loop with respect to the magnetic fields applied at 2 K. %
Moreover, the width of the hysteresis loop roughly agrees with that of the $M$-$H$ curves shown in Fig. \ref{mag}(c). %
Although the agreement between these hysteresis loops is not perfect, this is probably because we measured the remanent electric polarization at $H=0$ after sweeping magnetic field up to a certain value, while we measured the magnetization with varying $H$. %
From these results, we concluded that $P_c$ is coupled with the weak ferromagnetic moment, revealing that the $\sigma$-induced ferroelectricity originates from the spin order. %

\begin{figure}[t]
\begin{center}
	\includegraphics[clip,keepaspectratio,width=8cm]{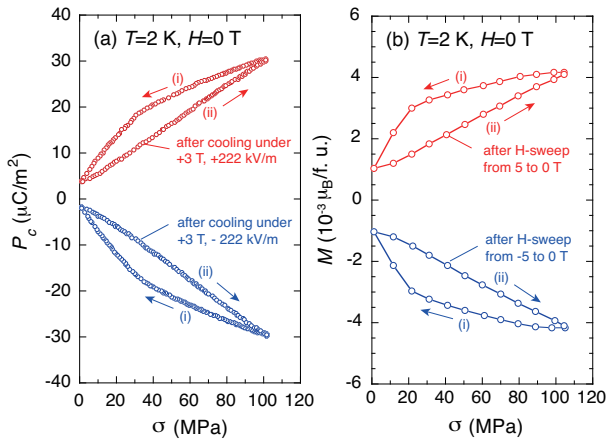}
	\caption{(Color online) (a) Piezoelectric responses at $T=2$ K in zero magnetic field measured after cooling under  $\sigma=101$ MPa, $H=3$ T and $E_p\pm 222$ kV/m. %
	(b) Piezomagnetic responses at $T=2$ K in zero magnetic field measured after $H$-sweeping from 0 to $\pm 5$ T and then to 0 T under  $\sigma=101$ MPa.}
	\label{piezo}
\end{center}
\end{figure}

Figure \ref{piezo}(a) shows piezoelectric responses at 2 K in zero magnetic field. %
Before the measurements, the sample was cooled under $\sigma=101$ MPa, $H=3$ T and $E_p\pm 222$ kV/m. %
We removed $H$ and $E_p$ at 2 K, and then measured the displacement current with varying $\sigma$ from 101 to 1 and then back to 101 MPa. %
The absolute values of $P_c$ were determined by the subsequent pyroelectric measurements on heating. %
We observed that the $\sigma$-induced $P_c$ changes in accordance with the varying $\sigma$, and that the polarity of the $P_c$ changes depending on $E_p$ applied on cooling. %

Similarly, we also confirmed the piezomagnetic nature as shown in Fig. \ref{piezo}(b). %
We swept $H$ from 0 to 5 (or $-5$), and again to 0 T under $\sigma$, and subsequently measured the spontaneous magnetization at $H=0$ with varying $\sigma$ from 101 to 1, and again to 101 MPa. %
We observed that the $\sigma$-induced ferromagnetic moment also shows a similar behavior to that of the $\sigma$-induced $P_c$. %
Note that the hystereses in these piezoelectric and piezomagnetic responses are probably due to the imperfect reduction of the stress; because of the friction between the zirconia pistons and anvils used in the pressure cell, the actual value of $\sigma$ could not follow the nominal load value in particular in the $\sigma$-decreasing processes. %

Finally, we discuss the microscopic picture of the uniaxial stress effects on the magnetic orders. %
As mentioned above, the disappearance of the weak ferromagnetism at $T_{\rm SR}$ is related to the rotation of the $G$ component of Fe spins \cite{DFO_Gorodetsky_JAP}. %
The present experiments demonstrate that an application of $\sigma$ induces weak ferromagnetic moments in the AFM and Dy-ordered AFM phases. %
From these results, we suggest that the application of $\sigma$ affects in-plane magnetic anisotropy of the Fe spins, so that they are slightly rotated from their original position. %
The rotation of the Fe spins can yield a small $G_x$ component, which induces the spontaneous electric polarization in the Dy-ordered AFM phase via the magnetostriction mechanism, in the same manner as in the field-induced multiferroic phase \cite{Tokunaga_DFO_PRL,NPhys_Tokunaga_RFO}. %
By comparing the values of $M$ and $P_c$ in the Dy-ordered AFM phase under $\sigma$ with those in the magnetic-field-induced multiferroic phase, the rotation angle is roughly estimated to be $1\sim 2$ degrees at $\sigma=151$ MPa. %

In summary, we have investigated uniaxial-stress effects on magnetic and dielectric properties of DyFeO$_3$, which has a centrosymmetric crystal structure. %
By means of magnetization and pyroelectric measurements, we revealed that an application of $\sigma$ induces both the weak ferromagnetism and ferroelectricity in the Dy-ordered AFM phase. %
We also demonstrated that the magnetic-field-induced reversal of the weak ferromagnetic moment is accompanied by the flip of the electric polarization, revealing that the two $\sigma$-induced ferroic orders are coupled with each other. %
As for the microscopic mechanism of the uniaxial-stress effects, we suggest that the application of $\sigma$ results in the change in magnetic anisotropy, which leads to a rotation of the Fe spins. %
This rotation yields the small $G_x$ component of the Fe spins, which is relevant to the emergence of the weak ferromagnetism and also induces ferroelectricity through the magnetostriction mechanism. %
This work provides a new guiding principle for designing spin-driven multiferroicity. %
By applying anisotropic stress on an antiferromagnet, one can remove a number of symmetry operations in the system. %
This may lead to magnetically and electrically polarized states, in which the degree of the polarization can be tuned by the applied uniaxial stress. %
This guideline is also applicable for thin film samples \cite{DFO_film_APL,Nakamura_YMO_film,BFO_film_Science}, in which structural mismatch between the film and substrate often results in significant lattice deformation. %
By utilizing the epitaxial strain from the substrate, one may possibly impart additional ferroic orders to the film. %

This work was supported by Grants-in-Aid for Young Scientists (A) (No. 25707032) and (B) (No. 25800203) from JSPS, Japan.  
The images of the crystal and magnetic structures in this paper were depicted using the software VESTA \cite{VESTA} developed by K. Momma. %

\bibliography{main}

\end{document}